\documentclass[a4paper,fleqn,usenatbib]{mnras}

\usepackage[T1]{fontenc}
\usepackage{ae,aecompl}

\usepackage{graphicx}	
\usepackage{amsmath}	
\usepackage{amssymb}	

\title[Synchrotron emission modelling of AGN jets with RHD simulations]{Using synchrotron emission modelling of relativistic hydrodynamic jet simulations to study the FR I/FR II dichotomy of Active Galactic nuclei radio jets}

\author[I.P. van der Westhuizen et al.]{
I.P. van der Westhuizen,$^{1}$\thanks{E-mail: vanderwesthuizenip@ufs.ac.za}
B. van Soelen,$^{1}$
P.J. Meintjes,$^{1}$
and J.H. Beall$^{2,3}$\\
$^{1}$Department of Physics, University of the Free State, 205 Nelson Mandela Drive, Bloemfontein, 9301, South Africa \\
$^{2}$ St. John’s College, Annapolis, USA\\
$^{3}$ Space Sciences Division, Naval Research Laboratory, Washington, DC, USA}

\date{Accepted XXX. Received YYY; in original form ZZZ}

\pubyear{2015}

\begin{document}
\label{firstpage}
\pagerange{\pageref{firstpage}--\pageref{lastpage}}
\maketitle

\begin{abstract}
In this paper three dimensional relativistic hydrodynamic simulations of AGN jets are presented to investigate the FR I/FR II dichotomy. Three simulations are presented which illustrates the difference in morphology for high/low Lorentz factor injection as well as a stratified background medium. Lorentz factors of 10 and 1.0014 were used for the high and low Lorentz factor cases respectively. The hydrodynamic simulations show a division in the morphology of jets based on their initial injection luminosity. An additional simulation was set up to investigate the evolution of the low Lorentz factor jet if the mass injection was lowered after a certain time. A synchrotron emission model was applied to these simulations to reproduce intensity maps at radio frequencies (1.5 GHz) which were compared to the observed emission structures of FR I/FR II radio galaxies. The effect of Doppler boosting on the intensity maps was also investigated for different polar angles.  The intensity maps of both the high and low Lorentz factor cases reproduced emission structures that resemble those of FR II type radio galaxies with a dominant cocoon region containing time dependent hot spots and filaments. An FR I like structure was, however, produced for the low Lorentz factor case if the mass injection rate was lowered after a set time period.      
\end{abstract}

\begin{keywords}
hydrodynamics -- radiation mechanisms: non-thermal -- methods: numerical -- galaxies: jets
\end{keywords}



\section{Introduction}

Radio-loud active galactic nuclei (AGN) produce relativistic jet structures that can span hundreds of kiloparsec. These jets have complex time dependent morphologies, however, on the largest scales their general morphology is classified into two Fanaroff-Riley types \citep{1974MNRAS.167P..31F}. Fanaroff-Riley class I (FR I) sources have low luminosity jets where the radio luminosity decreases with distance from the core, while Fanaroff-Riley class II (FR II) sources show high luminosity jets in which the jets increase with brightness further from the host galaxy \citep{1974MNRAS.167P..31F}. In addition to these two classes a third class of FR 0 sources has recently been suggested, which are core dominated and lack the extended jet structures observed in the FR I and FR II classes \citep{2015A&A...576A..38B}. 

In order to develop a complete understanding of the intrinsic and extrinsic differences between the different classes of AGN the mechanisms responsible for the production of different jet morphologies must be investigated. Many models have proposed elements which contribute to the dichotomy of these classes. One prominent model developed by \citet{1974MNRAS.169..395B} proposed a ``twin jet" for Cyg A type sources in which the hot spots and radio lobes are powered by a relativistic beam. The kinetic energy of the bulk motion in the beam is transformed into internal energy at the interface between the jet and ambient medium, referred to as the working surface. This conversion of energy drives the formation of the FR II type emission. 

The conditions required for FR I type radio jets are not well understood, with most models constraining the dichotomy through injection luminosity and the ambient medium density profile \citep[see e.g.][]{1991MNRAS.250..581F, 1997MNRAS.286..215K}. Another condition that may be important for the production of FR I type radio jets is a decelerating beam \cite[see e.g.][]{2002MNRAS.336.1161L}. To investigate possible mechanisms through which the beam deceleration can occur, many studies have turned to hydrodynamic simulations. For example, \cite{2007MNRAS.382..526P} created a hydrodynamic simulation of the FR I source 3C 31 based on the study by \cite{2002MNRAS.336.1161L}, and showed that the re-collimation shocks in a light jet trigger non-linear perturbations that can lead to the mixing of external medium with the jet beam. This entrainment of mass from the ambient medium in the beam of the jet is the most probable cause for the deceleration of the beam. The results presented in \cite{2007MNRAS.382..526P} were also consistent with $x$-ray observations of FR I sources (see e.g. Croston et al., 2007; Kraft et al., 2003). There are several additional mechanisms which can promote the process of ambient mass entrainment. \cite{2008A&A...488..795R} suggested that shear instabilities may be the main mechanism for entrainment. In this mechanism, the jet to ambient density ratio plays a crucial role with larger deceleration occurring in light jets. Recently \cite{2016A&A...596A..12M} showed that hydrodynamic simulations of decelerated beams reproduce jet structures with physical characteristics that are associated with FR I type AGN. This study simulated the dynamics of beams which have already been decelerated to a non-relativistic regime and estimated an injection power of $10^{43}$ erg.s$^{-1}$ as a division in the formation of FR I and FR II type jets. 

While the aforementioned studies focused on the simulated physical properties of the jet, few have modelled the resulting synchrotron emission responsible for Fanaroff-Riley classes. When investigating the morphology of AGN jets it remains important to not only consider the formation of physical structures in hydrodynamical simulations, but also their associated emission. An example of this was shown by \cite{1997ApJ...482L..33G} which modelled the synchrotron emission of an $2.5$D axis symmetric model to show that simulations can reproduce superluminal emission components. Their synchrotron model was based on the assumptions that there is a proportionality between the number and energy density of the thermal fluid and non-thermal synchrotron emitting electrons. Studies such as \cite{2009ApJ...696.1142M} and \cite{1999ApJ...512..105J} took a self-consistent approach to the modelling of synchrotron emission, where the evolution of non-thermal particles is simulated in addition to the thermal fluid in $2.5$D axis symmetric simulation to model the emission properties for smaller scale parsec jets.

In this paper we combine synchrotron emission modelling with several 3D numerical hydrodynamic jet simulations to investigate how certain physical structures relate to the production of FR I/FR II type radio jets. The numerical simulations that are presented focus on high and low Lorentz factors to specifically investigate the influence of a decelerated beam on the emission. An emission model, in the form of a post-processing code, is applied to the hydrodynamic simulations to generate approximate intensity maps of the synchrotron radiation produced by the environment. These maps are used to relate the simulations to the FR I/FR II dichotomy. The hydrodynamic simulations and setups are discussed in section 2. Section 3 describes the implementation of the emission modelling, while section 4 contains our results and is followed by conclusions in section 5.

\section{Hydrodynamic Simulations}

The hydrodynamic simulations were designed and compiled using the {\sc pluto} software,\footnote{http://plutocode.ph.unito.it/} which uses upwind high resolution shock capturing algorithms to solve hydrodynamic conservation equations and evolve them with time  \citep{2007ApJS..170..228M}. The code uses a grid based algorithm that consists of a set of defined cells in a mesh grid with assigned properties that adhere to the fluid dynamic conservation laws. The relativistic hydrodynamic (RHD) module of the code was used to allow for a relativistic bulk motion in the jet. This approach to the numerical simulations is valid under the assumption that the jet is kinetically dominated, with the magnetic field having negligible effects on the dynamic morphology of the jet. Such a regime may occur at kiloparsec scale distances from the host galaxy and therefore it is also assumed that the effects of gravitation on the jets will be negligible. Only the numerical viscosity produced by the solvers are present in the simulations and the effects of radiative cooling were also neglected. 

The simulations' environments consisted of a three dimensional Cartesian domain spanning $12.8\times6.4\times6.4$ kpc. The jet medium was injected through a circular nozzle, of radius $100$ pc, on the lower $z$ boundary. The simulations consisted of light jets with jet material less dense than the ambient medium. Studies such as \citet{1997ApJ...479..151M} and \citet{2008A&A...488..795R} have shown that the propagation efficiency of the jet also becomes less with a smaller density ratio. \citet{2008A&A...488..795R} found a density ratio of $10^{-3}$ may produce FRI like morphologies, therefore, our cases was chosen with a similar density ratio. The injection nozzle was set up with a profile to ensure a smooth transition between the jet and ambient medium. An initial pressure equilibrium between the nozzle and the external medium was used to ensure that the jets were initially collimated. Reflective boundary conditions were chosen for the lower $z$ boundary to simulate the presence of a galaxy, while all other boundary conditions were set-up as outflow boundaries.

Four simulations are presented for this study. The first two simulations were constructed to investigate the effects of a decelerated beam and were therefore setup with a large and a small Lorentz factor respectively. The kinetic luminosity was calculated for each simulations to ensure that they were separated by the $10^{43}$ erg.s$^{-1}$ division \citep[see ][]{2002MNRAS.331..615S}. The properties assigned to the ambient and jet material are listed in Table \ref{tab:parameters}. The parameters for the high Lorentz factor case (Case A) was chosen based on previous studies \citep[e.g. Case D in][]{2008A&A...488..795R} which showed a strong terminal shock with injection luminosity comparable to $10^{43}$ erg.s$^{-1}$. The lower Lorentz factor simulation was chosen based on the results shown in \citet{2016A&A...596A..12M}. The background medium of these simulations were stratified to simulate the density profile of the inter galactic medium, with the density decreasing as,
\begin{equation}
\rho(r)=\frac{\rho_b}{1+(\frac{r}{40})^2},
\end{equation}
where $\rho_b$ is the ambient density at the injection point and $r$ is the distance in beam radii from the injection point \citep[see e.g.][]{1991MNRAS.250..581F, 1997MNRAS.286..215K, 2016A&A...596A..12M}. A base density of $\rho_b=10^{-24}$ g.cm$^{-3}$ was used for the ambient medium in all of the simulations \citep{2008A&A...486..119B}. The third simulation was set up with a uniform background medium and high Lorentz factor injection to investigate the effects of the stratified background medium. In all three cases a uniform pressure was initially assigned to the background medium. 

A fourth simulation was set up to investigate how a change in the injection rate effects the morphology. In this simulation the jet was initially injected with the same parameters as in case B. After the jet had evolved to a stable point the injection density was lowered by a factor of 5, while the Lorentz factor and Mach number were kept constant. This was done to lower the injection luminously of the jet.

The parameters that were chosen are supported by observational studies. For FR II type galaxies \citet{2006ARA&A..44..463H} found Lorentz factors ranging between $5\sim40$. There is no definitive separation between the initial Lorentz factors of FRI and FRII class jets, however FRI jets show much faster deceleration on kiloparsec scales. We consider an extreme case of a decelerated beam with a Lorentz factor 1.0014. The choice of Mach numbers was based on the temperature ranges for the ambient medium provided by studies of early type galaxy coronas \citep[see e.g][]{2008A&A...486..119B, 2013MNRAS.433.2259P}. The ambient temperature of the relativistic cases (A and C) at the base of the simulation then correspond to $0.062$ keV. Inside the jet at the injection point the temperature is $620$ keV for cases A and C. For the non-relativistic case the temperature of the ambient medium at the base of the simulation was calculated as $0.098$ keV, while the temperature inside the beam is $98$ keV.
  
The numerical algorithm that was used to evolve the properties of each cell with time consisted of parabolic spacial interpolation, the HLL Riemann solver and characteristic tracing time stepping  \citep{2005ApJS..160..199M}. The Taub-Matthews equation of state was used to describe the fluid \citep{2007MNRAS.378.1118M}. The simulations were run at the University of the Free State's High Performance Cluster (HPC) unit and were evolved until the jet structure reached the edge of the computational domain. 


\begin{table*}
	\centering
	\caption{Parameters used in the set up of the initial conditions.}
	\label{tab:parameters}
	\begin{tabular}{lcccccc} 
	\hline
Case & Dimensions & Resolution & Lorentz factor & Density ratio & Mach number & Kinetic luminosity \\ \hline
A & $512\times512\times1024$ & $1.25$ & $10$ & $10^{-4}$ & $30$ & $10^{45}$ erg.s$^{-1}$\\
B & $512\times512\times1024$ & $1.25$ & $1.0014$ & $10^{-3}$ & $4$ & $10^{42}$ erg.s$^{-1}$\\ 
C & $256\times256\times512$ & $2.5$ & $10$ & $10^{-4}$ & $30$ & $10^{45}$ erg.s$^{-1}$\\ 
D & $512\times512\times1024$ & $1.25$ & $1.0014$ & $5\times10^{-4}$ & $4$ & $10^{41}$ erg.s$^{-1}$\\ \hline
	\end{tabular}
\end{table*}

\section{Synchrotron emission}

To determine whether the physical structures of the simulations agree with the FR I/FR II dichotomy a post-processing code was developed to estimate the synchrotron emission. In this code the synchrotron emission can be calculated for arbitrary viewing angles and frequencies, and it produces 2D intensity maps in the observer frame. Geometric and relativistic effects are taken into account to correct for effects such as light travel time and Doppler boosted emission, however, cosmological redshifts are neglected. In our estimates the assumption was made that the synchrotron radiation was predominantly emitted by non-thermal electrons with a single power-law particle distribution, given by,
\begin{equation}
n^\prime_e(\gamma^\prime)=n^\prime_0 \gamma^{\prime-p}.
\label{eq:powerlaw}
\end{equation} 
Here the primed terms indicate quantities in the co-moving frame of the fluid (while the unprimed quantities are in the galactic stationary frame), $\gamma^\prime$ is the Lorentz factor, p is the power-law index and $n^\prime_0$ is the normalization factor, determined by \citep{1995ApJ...449L..19G}
\begin{equation}
n^\prime_0=\left( \frac{e^\prime(p-2)}{1-C_E^{^\prime 2-p}} \right)^{p-1} \left( \frac{1-C_E^{^\prime 1-p}}{\frac{\rho^\prime}{m_p} (p-1)} \right)^{p-2}.
\label{eq:normalization} 
\end{equation}
Here $m_p$ is the proton mass, $e^\prime$ is the internal energy density of the fluid and $C_E^\prime$ is the ratio of the maximum and minimum energies. For all calculations we have adopted $p=2.2$ and $C_E^\prime=10^3$. 

To construct artificial synchrotron intensity maps of the three dimensional simulations we first determined the radiation emitted and absorbed by each cell. The synchrotron emissivity is calculated by integrating the power radiated by a single radiating particle, $P^\prime_{\nu^\prime}(\gamma^\prime)$, over the particle spectrum $n^{\prime}_e(\gamma^\prime)$ of the medium \citep{rybicki1979radiative},
\begin{equation}
j^\prime_{\nu^\prime}  =\frac{1}{4\pi}\int n^{\prime}_e(\gamma^\prime) P^\prime_{\nu^\prime} (\gamma^\prime) \text{d} \gamma^\prime\text{.}
\label{eq:emcoeff}
\end{equation}

Since calculating the full expression requires a numerical integration, it will be computationally expensive to perform this for all $\sim10^8$ cells in the simulation. To compensate for this the synchrotron radiative power was approximated using an analytical $\delta$-function model \citep{dermer2009high, boettcher2012relativistic}. The $\delta$-function approximates the synchrotron power emitted by a particle of energy $\gamma^\prime$ assuming that the particle only radiates at a critical frequency $\nu^{\prime}_c$, given by,
\begin{equation}
\nu^\prime_c=\frac{3qB^\prime}{4\pi mc} \gamma^{\prime 2}\text{.}
\label{eq:nu_c}
\end{equation}
where $q$ is the charge of the radiating particle, $m$ is the rest mass of the radiating particle, $B^\prime=\sqrt{8\pi u_B}$ is the magnetic field in the co-moving frame and $c$ is the speed of light. The expression for the radiative power of the $\delta$-function approximation is given by,
\begin{equation}
P^{\prime\delta}_{\nu^\prime} (\gamma^\prime)=\frac{32\pi}{9} \left( \frac{q^2}{mc^2} \right) ^2 u_B^\prime \beta^{\prime 2} c \gamma^{\prime 2} \delta(\nu^\prime-\nu^\prime_c)\text{,}
\label{eq:delta} 
\end{equation}
where $\nu^\prime$ is the frequency of emission and $u_B^\prime$ is the magnetic energy density in the co-moving frame. 

Combining the previous equations \ref{eq:emcoeff}-\ref{eq:delta} and integrating equation \ref{eq:emcoeff} we obtain an expression for the emissivity,
\begin{equation}
j^\prime_{\nu^\prime}=\frac{4}{9} \left( \frac{q^2}{mc^2} \right) ^2 u_B^\prime \nu^{\prime\frac{1}{2}} \nu_0^{\prime-\frac{3}{2}} \beta^\prime 2 c n \left( \sqrt{\frac{\nu^\prime}{\nu^\prime_0}} \right)\text{,}
\end{equation}
where,
\begin{equation}
\nu^\prime_0=\frac{3qB^\prime}{4\pi mc}\text{.}
\end{equation}

A similar analysis is applied to the absorption coefficient,
\begin{equation}
\alpha^\prime_{\nu^\prime} = -\frac{1}{8\pi m \nu^{\prime 2}} \int P^{\prime\delta}_{\nu^\prime} (\gamma^\prime) \gamma ^{\prime 2} \frac{\partial}{\partial \gamma^\prime} \left( \frac{n(\gamma^\prime)}{\gamma ^{\prime 2}} \right) \text{d} \gamma^\prime \text{,}
\end{equation}
which leads to the $\delta$-expression,
\begin{equation}
\alpha^\prime_{\nu^\prime}=\frac{2}{9}\frac{p+2}{m\nu^{\prime 2}} \left( \frac{q^2}{mc^2} \right) ^2 u_B^\prime \nu_0^{^\prime-1} \beta^{\prime 2} c n \left(\sqrt{\frac{\nu^\prime}{\nu^\prime_0}}\right)\text.
\end{equation}
Considering that RHD simulations were used in this study, the magnetic field required to estimate the synchrotron emission could not be directly extracted from the RHD simulation. Current models for the formation of AGN jets suggest that strong toroidal magnetic fields collimate and accelerate these jets on sub-parsec to parsec scales, however, on the distance scales simulated in this study most of the magnetic energy density has been converted to the kinetic energy of the bulk flow \citep[see e.g ][]{2005ApJ...625...72S}. In order to obtain an estimate for the magnetic field it was assumed that a tangled field was present in the fluid, with the magnetic energy density ($u_B^\prime$) of each cell equal to a constant fraction of the internal energy density ($e^{\prime}$),
\begin{equation}
u_B^\prime=\epsilon_B e^\prime.
\end{equation}  
A fraction of $\epsilon_B=10^{-3}$ was chosen based on the assumption that the magnetic field energy density is small in comparison to the internal energy of the fluid. This fraction corresponds to a magnetic field strength on the order of $~10^{-4}$ G at the injection nozzle. Similar values of $\epsilon_B$ have been used in the past to model the spectral energy distributions of emission regions in AGN \citep[see e.g.][]{1995ApJ...449L..19G, 2000ApJ...528L..85A, 2014MNRAS.438.1856R, 2016A&A...588A.101F, 2004A&A...418..947M}, with \cite{2010ApJ...711..445B} showing that a change in $\epsilon_B^\prime$ does not significantly alter the synchrotron peak of the spectrum.

In order to produce two dimensional intensity plots of the synchrotron radiation, the coefficients are integrated along a line of sight $s$. The change in intensity (in the observer reference frame) can be given as \citep{rybicki1979radiative},
\begin{equation}
\frac{dI_\nu}{ds}= j_\nu - \alpha_\nu I_\nu\text{.}
\end{equation}  
The synchrotron coefficients of each cell transform as,
\begin{equation}
j_\nu=\frac{j_{\nu ^\prime}^{\prime}}{(\Gamma[1-\beta \mu])^2},
\end{equation}
\begin{equation}
\alpha_\nu=\alpha_{\nu ^\prime}^{\prime}(\Gamma[1-\beta \mu])
\end{equation} 
where $\mu$ is the cosine of the angle between the observer and the velocity of the fluid in the galactic stationary frame and $\Gamma$ is the Lorentz factor of the bulk flow.

\section{Results}

\subsection{Hydrodynamic simulations}
\begin{figure*}
	\includegraphics[width=40pc]{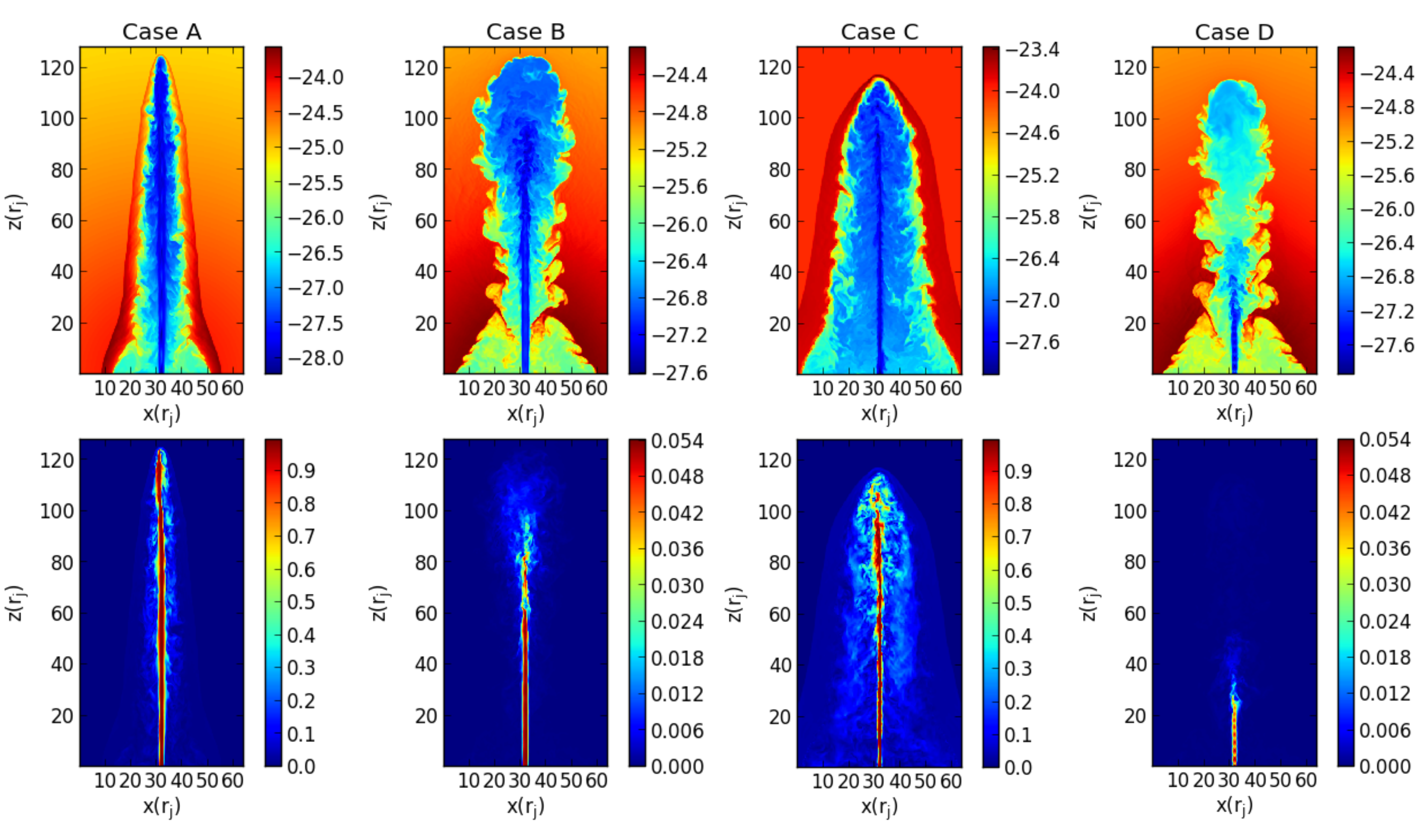}
    \caption{Density (top) and velocity (bottom) slices through the $xz$-plane of the simulations for Case A at a time of $9.5\times10^{4}$ yr, Case B at a time of $4.26\times10^{7}$ yr, Case C at a time of $1.6\times10^{5}$ yr and Case D at a time of $3.5\times10^{6}$ yr. Logarithmic scales are shown for the density plots in units of g.cm$^{-3}$, while the $x$ and $z$ axis are given in units of beam radius. The velocity is given in units of c.}
    \label{fig:2d_plots}
\end{figure*}

The numerical simulations were run with the set-up and parameters as discussed in section 2. Fig. \ref{fig:2d_plots} illustrates $xz$-slices of the density and velocity distributions for the four different cases. The formation of 3 distinct features are found within all four cases; first the outer uniform region that consists of the background material, second a collimated beam of jet material which forms the spine of the jet, and finally a mixture of jet and ambient medium surrounding the spine called the cocoon.

\begin{figure}
	\includegraphics[width=\columnwidth]{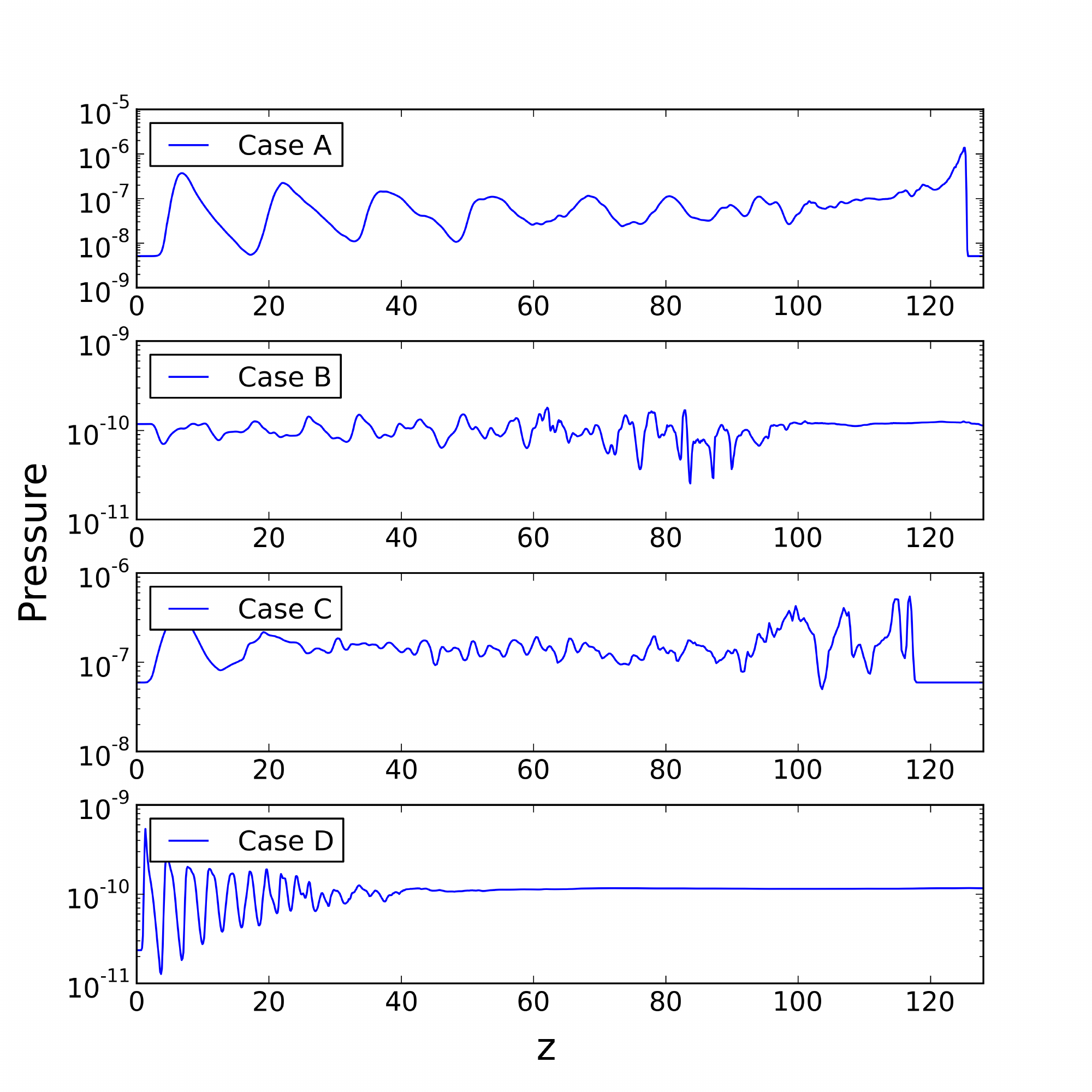}
    \caption{Pressure variation along the center of the environment in the $z$ direction for the time steps shown in Fig. \ref{fig:2d_plots},  in beam radii. Cases A and C show a termination shock between the jet and ambient medium.}
    \label{fig:Prs}
\end{figure}

In addition to the three features mentioned above cases A and C contain a terminal shock. Initially as material is injected into the environment, at supersonic speeds through the nozzle, it compresses the background material and forms a terminal shock surrounding the body of the jet. This terminal bow shock propagates through the ambient medium forming a region of turbulent high pressure material. This increase in pressure causes a pressure imbalance between the injected material and surrounding medium leading to the formation of Rayleigh-Taylor instabilities in the cocoon. In the high Lorentz factor simulations (Case A and C) the injection power is high enough for the jet propagation to remain supersonic, which generates a strong terminal shock. In the low Lorentz factor simulation the injection power is not enough to maintain the supersonic propagation of the jet and the terminal shock dissipates. The terminal shock is also shown in Fig. \ref{fig:Prs}, which plots the pressure along the central $z$-axis of the environment for each simulation. 

\begin{figure}
	\includegraphics[width=\columnwidth]{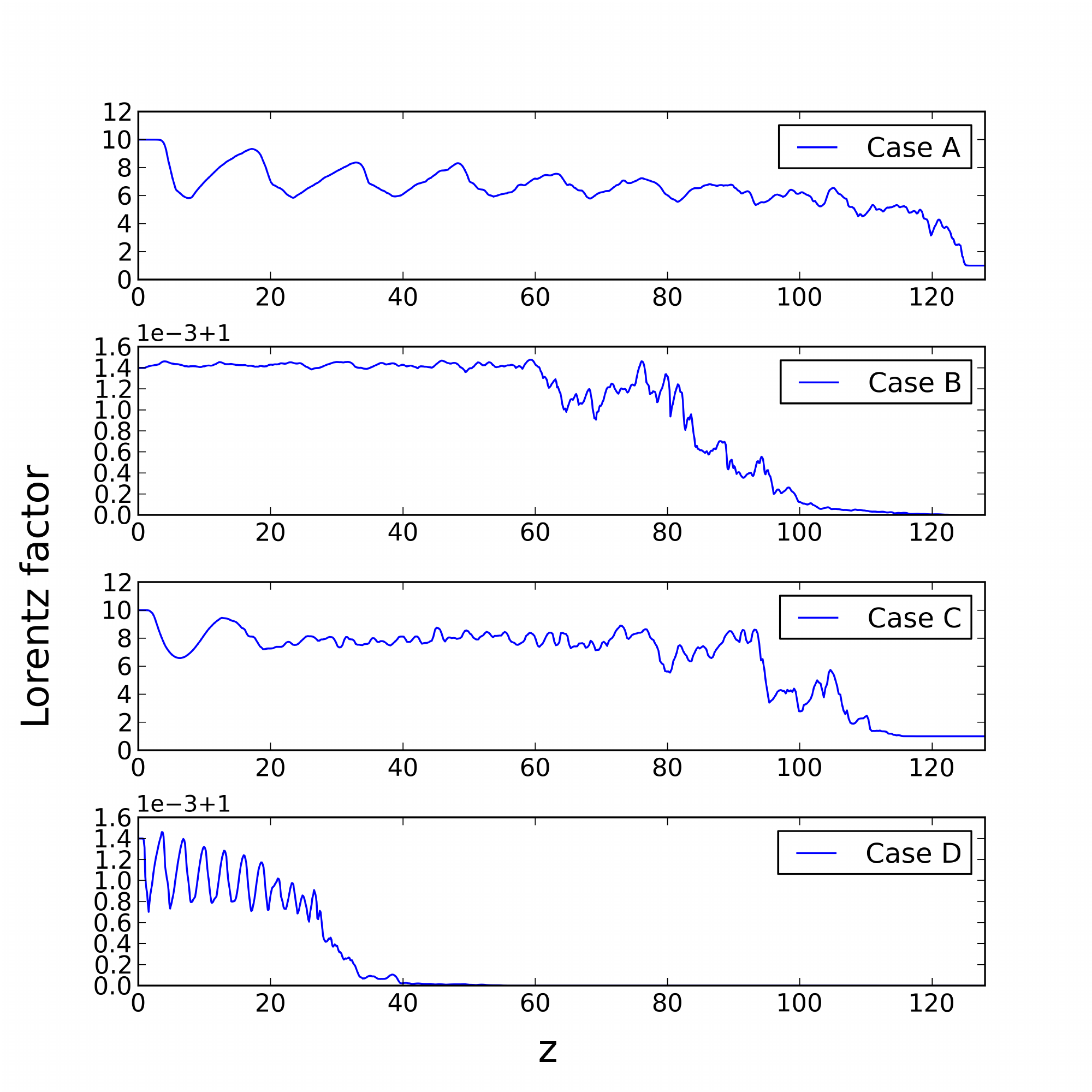}
    \caption{Plots of the maximum bulk Lorentz factor along the $z$-direction for all simulations at the time steps shown in Fig. \ref{fig:2d_plots}.}
    \label{fig:Lor}
\end{figure}

In the central region we obtain a collimated beam of jet material. In both cases A and C the beam is relativistic (see Fig. \ref{fig:Lor}, which plots the bulk Lorentz factor of the beam with $z$-distance) and shows little deceleration initially, with the bulk Lorentz factor decreasing from 10 to 5 along the beam for case A. Figs. \ref{fig:Prs} and \ref{fig:Lor} also show the formation of re-collimation shocks in the beam of cases A, C and D. The amplitude of these shocks declines with distance from the injection site. In case A, at a distance of 60 beam radii, the beam transitions from being re-collimation shock dominated to a turbulent regime. The turbulence forms due to hydrodynamic instabilities caused by the interaction between the beam of the jet and the surrounding cocoon region. A sharp deceleration in the bulk motion occurs at 108 beam radii at the working surface (the interface between the jet and ambient medium). A collimated beam also forms in case B despite the low Lorentz factor. Re-collimation shocks are, however, absent in this beam as it decelerates much faster. A slow moving region forms in the center of the beam between 50 and 60 beam radii.  

A cocoon formed, surrounding the jet beam in all of the simulations. This region is generated by a backflow of the jet material after it interacts with the shock front at the head of the jet. The structure of the cocoon (Fig. \ref{fig:2d_plots}) is divided into an inner region consisting of low density material and an outer, higher density, stationary region, that is generated by the mixing of ambient and cocoon material. The cocoon structure differs for all four cases, however, all of the cases show asymmetric, turbulent structures. When comparing the high and low Lorentz factors of case A and B we note that case A shows a narrow cocoon with the width of the cocoon generally decreasing with distance from the injection site, while case B shows a much broader cocoon with the diameter remaining almost constant with distance. This difference is due to the slower propagation of the jet through the ambient medium  in case B, allowing more material to accumulate in the cocoon. The effects of the stratified medium can be shown by comparing case A and C. Case C shows a less efficient propagation of the jet due to the denser ambient medium, which forms a broader cocoon. Case D shows a cocoon structure similar to that of case B, but with a much shorter beam.

The results shown by these simulations are in agreement to those of previous studies \citep[see e.g.][]{1997ApJ...479..151M, 2016A&A...596A..12M}. The presence of a bow shock surrounding the jet in simulations A and C have been associated with the production of the FR II type radio jets, while the absence thereof in simulations such as B and D has been suggested to resemble that of an FR I type radio jet. 

\subsection{Synchrotron emission}

\begin{figure*}
	\includegraphics[width=36pc]{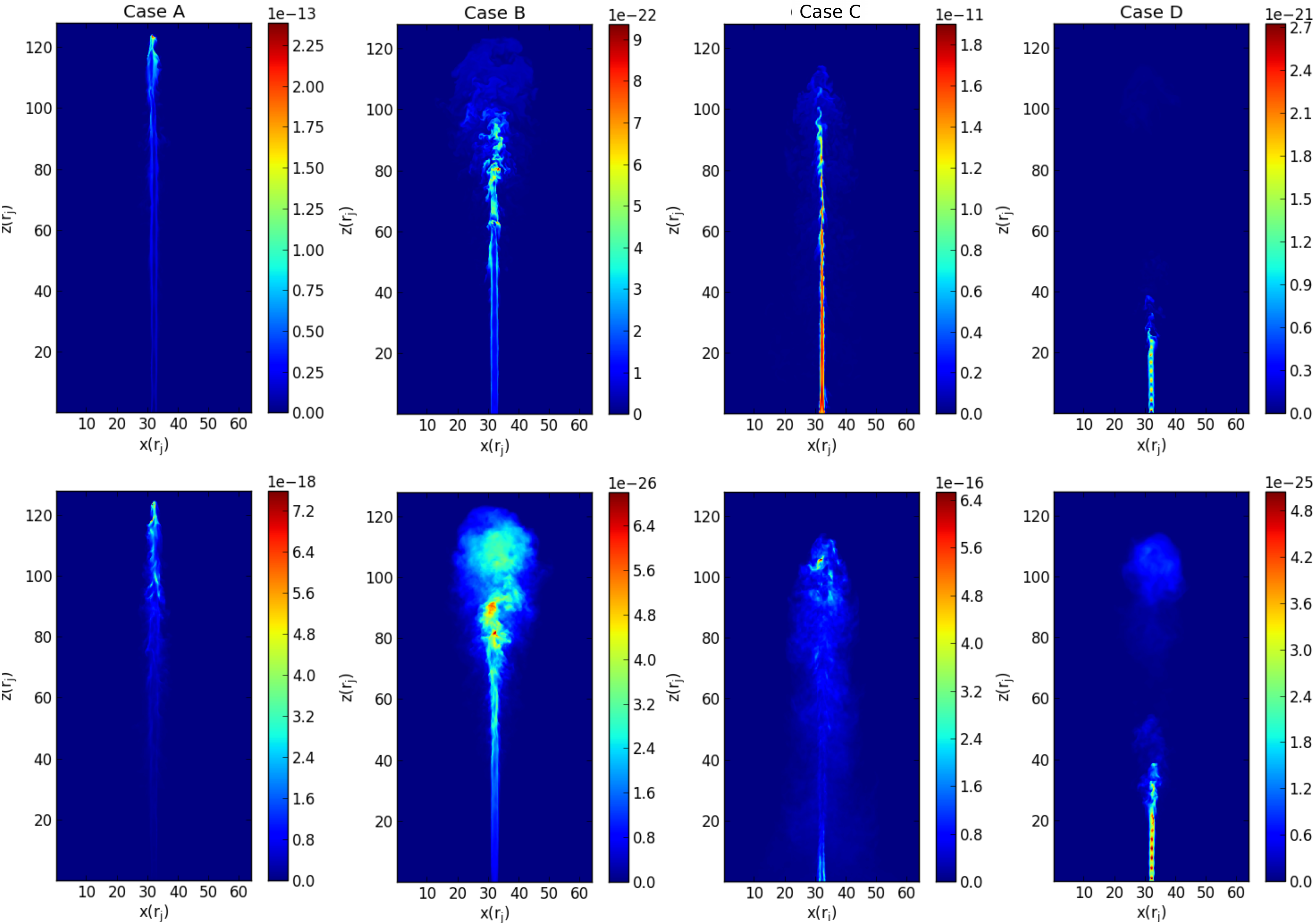}
    \caption{Slice of the emissivity in the co-moving frame (top) and Intensity maps (bottom), plotted for a viewing angle of $90\degr$ relative to the injection direction, for all four cases. Plots are scaled in arbitrary units. The time step for the emission at the centre of the frame corresponds to those shown in Fig. \ref{fig:2d_plots}.}
    \label{fig:2dss}
\end{figure*}

The post-processing emission code was applied to the hydrodynamic simulations and intensity maps were calculated at a frequency of $1500$ MHz. Fig. \ref{fig:2dss} shows the emissivity of the fluid in the co-moving reference frame ($xz$-slices) for all four cases (top) as well as the calculated intensity maps for a viewing angle of $90\degr$ relative to the jet-axis (bottom). In case A the highest emissivity occurs at the working surface between the jet and ambient medium. At this interface the kinetic energy is transformed into internal energy resulting in brighter emission. In both cases A and B we note the presence of a shear layer surrounding the beam, which has a higher emissivity (this is most prominent in case A). In case B the beam of the jet is decelerated before it reaches the head of the jet, which causes the highest emissivity to occur in the beam of the jet. In case C the beam of the jet also becomes less stable with distance, showing fragmented regions of brighter emissivity, however in contrast to case B the highest emissivity is obtained at small distances from the injection point. Case D shows the brightest emissivity in the jet of the beam with the brightest regions occurring in the re-collimation shocks. In all four cases it is shown the emissivity in the of the material in the jet beam is higher than that of the extended cocoon, however, the extended nature of the cocoon allows for brighter intensity in cases A, B and C when the emissivity is integrated along the line of sight. The absorption coefficients calculated for the simulations are proportional to the emissivity but remains low enough such that the environments are optically thin at 1.5 GHz.

The intensity maps show that cases A, B and C have the highest intensity in the cocoon regions. Cases A and C show hotspots corresponding to the working surface at the head of the jet, while the brightest region in case B occurs at a smaller z-distance similar to what is shown in the emissivity maps. These three cases also exhibit limb brightening, which is produced by a thin sheath layer which forms due to the interaction of the jet and ambient medium. 

The intensity maps shown for the first three cases are more consistent to that observed in FR II type radio galaxies than FR I, with the highest intensity emission occurring closer to the head of the jet. Case D on the other hand shows the highest intensity in the beam of the jet that fades with distance from the injection point. Re-collimation shocks in the beam of the jet are observed in the intensity maps as stationary emission regions. A diffuse region of emission is also observed at the upper edge of the cocoon, which was caused by the initial higher mass injection rate. The resulting structure of case D more closely resembles that of an FR I type radio jet.  

\begin{figure}
	\includegraphics[width=\columnwidth]{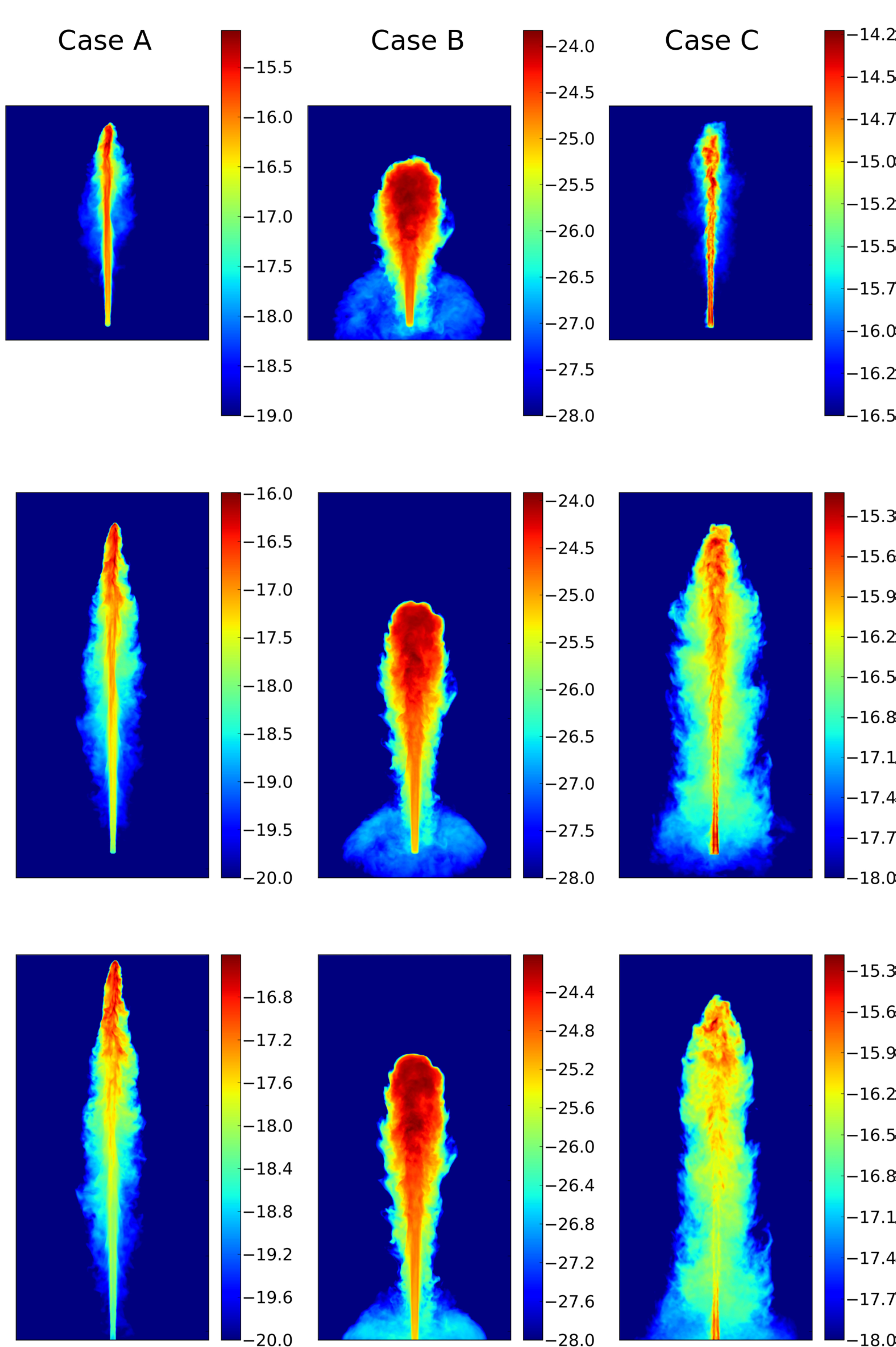}
    \caption{Intensity maps calculated for different polar angles relative to the jet  for cases A, B and C. The intensity maps are plotted for polar angles of $30\degr$ (top), $60\degr$ (middle) and $90\degr$ (bottom). Log scales are shown to emphasize the emission structure.}
    \label{fig:theta}
\end{figure}

One of the principles in the unified model of AGN is that the type of AGN is in part dependent on the inclination angle of the observer relative to the jet. This effect is caused by the Doppler boosting of emission due to the relativistic velocity of the jet plasma. To investigate how the emission structures in the simulated intensity maps are influenced by varying the viewing angle, intensity maps were calculated and plotted at different polar angles (defined as the angle between the injection direction and the $x$-axis) of $\theta=30\degr$ to $\theta=90\degr$ for cases A, B and C, shown in Fig. \ref{fig:theta}.  

The results show that changing the polar angle of the observation from an edge-on ($\theta=90\degr$) to a lower inclination angle causes the intensity of the relativistic beam to brighten significantly in the high Lorentz factor simulations (case A and C), while the intensity of the cocoon remains unchanged. This causes an overall increase in the intensity for lower polar angles. At low polar angles ($\sim\theta=30\degr$) the beam is the dominant emission region and the internal structure of the beam, which was obscured by the cocoon at high polar angles, can be clearly observed. In case C the beam structure does not show a uniform intensity, but a clumpy structure with individual emission components (most clearly illustrated at a polar angle of $\theta=30\degr$). Close to the injection site the re-collimation shocks discussed previously produces a small increase in brightness with symmetric distribution across the beam. At intermediate distances from the injection site, in the turbulent regime of the beam, the emission no longer exhibits a symmetric structure but appears deflected with small components breaking off from the main beam. At large distances where the beam has broken apart multiple, irregularly shaped emission regions are observed. An additional effect of the Doppler boosting observed in the results is that the edge brightening of the beam disappears as the polar angle is reduced. This confirms that the edge brightening is due to the sheath layer which has a lower Lorentz factor than the beam of the jet and subsequently is less Doppler boosted at low polar angles. The properties of the observed emission in such cases are therefore dependent on the orientation of the jet, since the dominant emission region changes. For the low Lorentz factor case (case B) the velocity of the jet fluid is not high enough for significant Doppler boosting to occur and, therefore, the structure of intensity map remain similar for all polar angles shown.  
 
\begin{figure}
	\begin{center}
	\includegraphics[width=14pc]{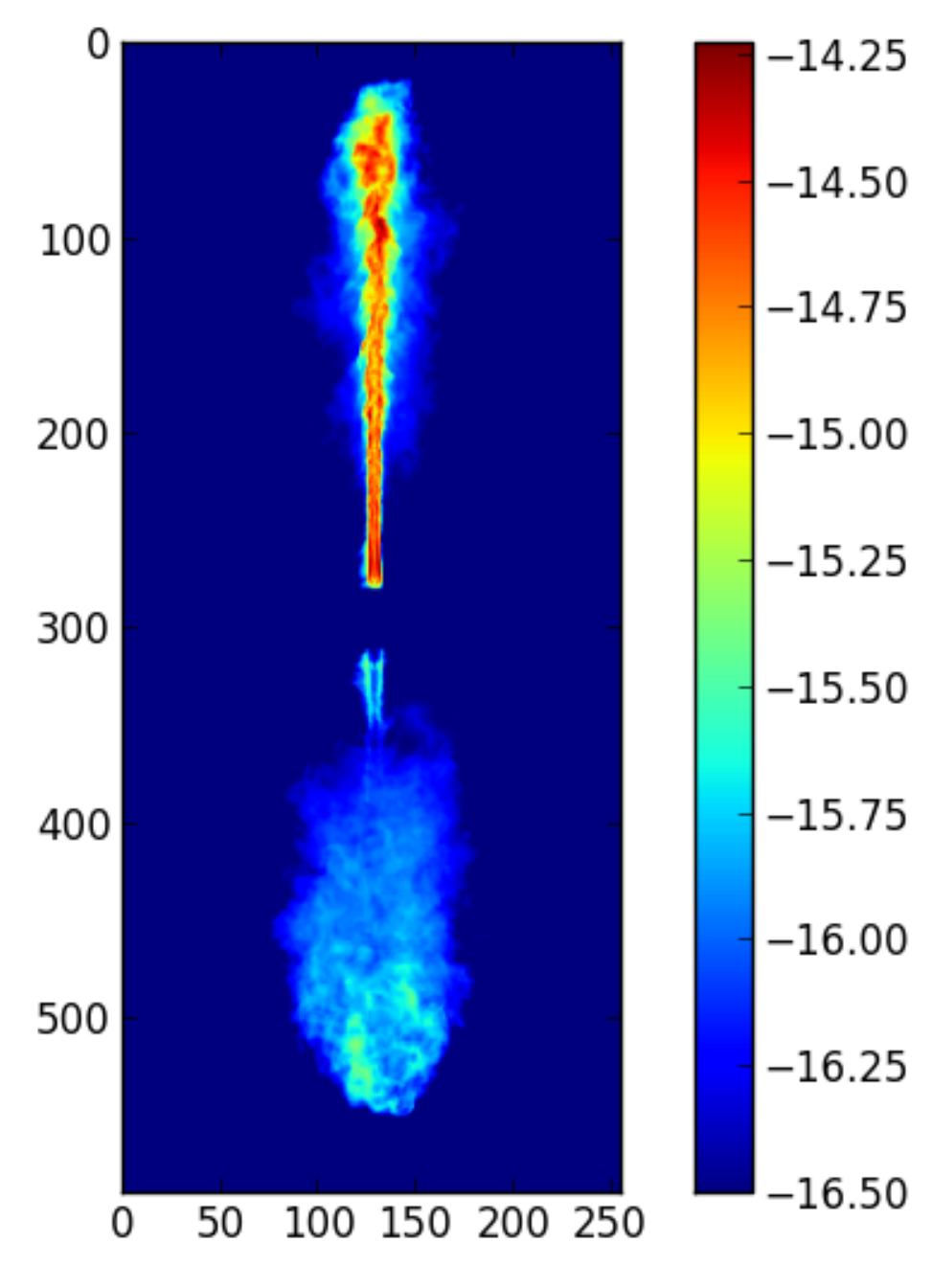}
    \caption{Intensity map calculated for identical bilateral jets with one facing the observer at an angle of $30\degr$ and the counter jet pointing away from the observer at a polar angle of $210\degr$.}
    \label{fig:comp}
    	\end{center}
\end{figure}

The resultant Doppler boosting shown in the intensity maps are consistent with observational studies, where FR {\sc II} radio galaxies with a high polar angle produce symmetric radio structures that are dominated by radio lobes, while low polar angle galaxies show asymmetric radio structures on either side of the galaxy with one component dominated by a relativistic beam and the other only by diffuse lobe emission. The effects of Doppler boosting on a galaxy with two symmetric radio jets are illustrated by constructing a composite image of two simulated intensity maps, calculated for polar angles of $\theta=30\degr$ and $\theta=210\degr$ (shown in Fig. \ref{fig:comp} for case C). Once again a similar large scale morphology is observed, with a prominent relativistic jet visible on the side which is Doppler boosted towards the observer. On the opposite side the emission from the cocoon region is dominant, while the beam of the jet is only visible at small distances due to limb brightening.  

\section{Discussion}
In this paper we show four different simulations of ideal relativistic jets produced using the PLUTO hydrodynamic code and the subsequent calculation of the expected synchrotron emission. Below follows a discussion of the results and the possible limitations arising from the assumptions and approximations that were made.

\subsection{Results}
All of the simulation results show the formation of collimated relativistic beams surrounded by a cocoon region of backflowing material. These results are in accordance with previous studies such as \cite{2008A&A...488..795R, 1997ApJ...479..151M, 2016A&A...596A..12M}. The results show that for a high bulk Lorentz factor ($\Gamma=10$) injection, the beam is initially dominated by re-collimation shocks which transitions to a turbulent regime further from the injection site. The beam remains relativistic until it reaches the working surface. This produces a strong shock at the interface between the jet and ambient medium and serves as a region for particle acceleration. For a low bulk Lorentz factor ($\Gamma=1.0014$) no re-collimation shocks were initially observed, and it was shown that the beam decelerates faster and breaks apart before reaching the head of the jet. These results show a division in the morphology of jets based on their initial injection velocity and therefore injection luminosity.
 
The effect of a stratified background medium was investigated and it was shown that a gradual decrease in density leads to an improved propagation efficiency of the jet, allowing for the formation of a longer jet over similar time scales. The propagation efficiency also influences the formation of the cocoon where a less efficient propagation leads to a larger cocoon diameter. In the high Lorentz factor case this led to the formation of a narrow cocoon region, which, in turn does not reproduce the typical radio lobes which have been observed with FR II sources. The cocoon diameter for this simulation will increase with time, however, to investigate the evolution of the system over larger time scales a larger domain is required. Observations of FR II AGN jets have shown that they can span up to 100 times the scales which have been simulated. Such a large simulation domain would be too computationally expensive to run with our currently available hardware.

We also investigated the effect of lowering the mass injection rate. It was shown that a collimated beam is maintained, however, the beam decelerates faster and breaks apart at smaller distances. The beam also developed re-collimation shocks which were caused by a pressure imbalance between the lower density jet material and the ambient medium. The cocoon initially formed from the higher density injection continues to propagate outwards, at a slower rate than the constant injection.   

Beyond the differences shown in the structure of the jet from the hydrodynamic simulations, the differences between FR I and FRII jets depends on the observed emission. We modelled the synchrotron emission from the whole simulated environment in order to investigate how this compares to observed radio jet structures. Estimates of the synchrotron emission generated for cases A and C show large scale similarities to FR II type radio jets with a relativistic beam and brighter lobe-like structures at 1500 MHz. Hot spots were shown to occur at the positions of the working surface which is consistent with the kinetic energy of the bulk flow being converted to internal energy. This is consistent with previous discussion of the hydrodynamic structure, and the majority of the emission originates from the region near the working surface. 

While the hydrodynamic simulations in case B show a structure which is closer to an FR I type jet,  with no strong shock, the simulated emissions does not follow this structure. The intensity map calculated for case B shows that even though the highest emissivity occurs in the beam and not in the head of the jet the extended cocoon still produces a higher intensity than the beam. This produces a simulated intensity map which is more consistent with an FR II type jet rather than an FR I. This may indicate that it is possible to reproduce FR II like structures with a decelerated beam and the absence of a shock front between the jet and ambient medium. 

There are several factors that may cause the lower Lorentz factor jet to be inconsistent with FR I structure. First, it is possible that the jet resulting from the simulation is in a transitional state where the expanding cocoon will cool with time to form an FR I type jet. However, when comparing the intensity map calculated at a time of $\sim10^7$ yr to that of an earlier time step $\sim10^5$ yr it showed a similar structure, which would suggest a stable state. Second the injection in the simulation is constant, and limited turbulence occurs in the initial beam. This may point to variability in the injection as a requirement for producing turbulence and consequently higher intensity from the beam.  

In case D we showed that if the initial mass injection rate (used in case B) is lowered, the resulting simulated intensity maps reproduces a FR I type jet, with a decrease in the brightness from the injection nozzle. The lower pressure of the injection leads to the formation of re-collimation shocks, which converts the kinetic energy of the bulk flow to internal energy and decelerates the beam. This produces stationary bright components in the jet beam at the location of these shocks. This shows that mechanism by which the beam is decelerated is important. This may indicate that variability of the injection is an important consideration, since the lower injection luminosity alone (case B) results in a simulated structure which is more consistent with a FR II and not FR I.  This result could also suggest that there are evolutionary ties between the FR types, since is shown that an FR II like morphology can evolve into an FR I like morphology if the mass ejection declines. 

\subsection{Possible limitations}
In this paper we have made a number of assumptions in producing the synchrotron emission maps. In this section we consider the possible limitations of these assumptions and how it would influence the results. First, the overarching assumption was that the fluid dynamic model of relativistic jets followed the model outlined by \citep{1974MNRAS.167P..31F}. In our simulations only RHD equations were considered assuming that the magnetic field strength is too small to have a significant impact on the overall morphology. There may occur local regions in which the kinetic energy of the bulk fluid becomes small, however we do not expect to find magnetic instabilities such as those shown, for example, in \cite{2010MNRAS.402....7M}. 

Second, the numerical method used to calculate the synchrotron emission assumes a single power-law spectrum for all non-thermal electrons with the emission normalized based on the internal energy density and number density of the fluid. This may be a reasonable assumption to reproduce intensity maps at a single frequency, however, it would not hold for estimating the shape of the spectral energy distribution. In addition, while this assumption produces maxima at the positions of shock fronts, it may lead to an over estimation of the cocoon's intensity, since the non-thermal electron spectrum will dynamically evolve with processes such as shock acceleration and radiative cooling. Along with this it was assumed that the energy radiated away by the non-thermal electrons is small in comparison to the bulk energy of the fluid and, therefore, the back-reaction between the emitting electrons and the fluid was not considered. 

Third, we should consider the assumption regarding a tangled magnetic field. While traditionally observational studies have used an equipartition relation to comment on magnetic field strength, this does not hold in general \citep[see e.g][]{2005AN....326..414B}. The assumption of proportionality between the magnetic field and the internal energy density (as used in the emission modelling) does not yield a uniform magnetic field since the energy density of the fluid evolves in each cell, however, it does maintain a constant relation to the thermal energy of the plasma. RMHD simulations such as those presented by \cite{2007A&A...466...93M} (1D shock collisions) and \cite{2005A&A...436..503L} (2.5D jet simulations) show variations in $\epsilon_B$ do occur. This may, in our simulations, lead to an over estimation of the magnetic field in regions such as shocks fronts. \cite{2007A&A...466...93M}, however, notes that the difference in their computed synchrotron emission for a constant $\epsilon_B$ and the $\epsilon_B$ obtained from RMHD simulations becomes insignificant for small values of $\epsilon_B<10^{-2}$. Similarly we expect that since the magnetic field energy density is much lower than the energy density of the radiating electrons the resulting deviation of a constant $\epsilon_B$ from the true ratio will have a small effect on the total integrated emission. The choice of a constant $\epsilon_B$ which would mostly effect local regions of change like shock fronts in the jet but should not significantly alter the overall morphology which was obtained.

The limiting factors mentioned cannot be implemented in the simulation presented in this study, should be investigated in future to determine their impact on the large scale emission morphology.

\section{Conclusion}

In this study the morphology of FR I and FR II AGN jets has been investigated by looking at different injections luminosities as well as stratification of the background medium and how this influences the overall physical structure of a jet. We found similar results to previous authors, which have pointed to a divide between FR I and FR II jets. However, since the actual observed structure of the jet depends on the synchrotron emission and not directly on the physical parameters calculated in RHD simulations we have expanded on this work to calculate an approximation of the observed synchrotron emission directly from the RHD simulations. While previous authors have looked at smaller regions inside jets, such as shock fronts, we have considered the full simulations and have therefore made approximation of the particle distribution and the magnetic field strength previously considered in the literature. We found for the four simulations: 
\begin{enumerate}
\item{The stratification of the background has a limited effect on the structure of the jet, though it allowed for a more efficient propagation.}
\item{The synchrotron intensity map estimations for the high luminosity jet follow the structure of the density profile, producing a FR II like morphology as expected.}
\item{For the low kinetic luminosity case, the synchrotron estimate, while showing a higher emissivity in the beam of the jet, the emission was dominated by the faint cocoon. We suggest that this is due to the lack of further deceleration mechanisms in the beam, perturbation in the injection luminosity or ordered magnetic field structure.}
\item{Lowering the injection density and pressure triggers standing shocks in the beam due to an over pressure environment, and causes the beam emission to become dominant. This suggests the nature of how the jet is decelerated is important for the formation of FR I type sources and that a constant low luminosity injection alone is insufficient to produced FR I structures.}
\item{We note that since a constant power-law index and ratio of equipartition is assumed, variations in the local parameters may introduce additional effects, that may produce some additional brightening of the beam. This should be investigated in future studies.}
\end{enumerate}

\section*{Acknowledgements}
The financial assistance of the National Research Foundation (NRF) towards this research is hereby acknowledged. This work is based on the research supported in part by the National Research Foundation of South Africa for the grants 87919 and 112673. Any opinion, finding and conclusion or recommendation expressed in this material is that of the authors and the NRF does not accept any liability in this regard. This research was in part supported by a UFS CRF grant. The numerical calculations were performed using the University of the Free State High Performance Computing Unit.




\bibliographystyle{mnras}
\bibliography{Ref} 

\begin{thebibliography}{}
\makeatletter
\relax
\def\mn@urlcharsother{\let\do\@makeother \do\$\do\&\do\#\do\^\do\_\do\%\do\~}
\def\mn@doi{\begingroup\mn@urlcharsother \@ifnextchar [ {\mn@doi@}
  {\mn@doi@[]}}
\def\mn@doi@[#1]#2{\def\@tempa{#1}\ifx\@tempa\@empty \href
  {http://dx.doi.org/#2} {doi:#2}\else \href {http://dx.doi.org/#2} {#1}\fi
  \endgroup}
\def\mn@eprint#1#2{\mn@eprint@#1:#2::\@nil}
\def\mn@eprint@arXiv#1{\href {http://arxiv.org/abs/#1} {{\tt arXiv:#1}}}
\def\mn@eprint@dblp#1{\href {http://dblp.uni-trier.de/rec/bibtex/#1.xml}
  {dblp:#1}}
\def\mn@eprint@#1:#2:#3:#4\@nil{\def\@tempa {#1}\def\@tempb {#2}\def\@tempc
  {#3}\ifx \@tempc \@empty \let \@tempc \@tempb \let \@tempb \@tempa \fi \ifx
  \@tempb \@empty \def\@tempb {arXiv}\fi \@ifundefined
  {mn@eprint@\@tempb}{\@tempb:\@tempc}{\expandafter \expandafter \csname
  mn@eprint@\@tempb\endcsname \expandafter{\@tempc}}}

\bibitem[\protect\citeauthoryear{{Aloy}, {G{\'o}mez}, {Ib{\'a}{\~n}ez},
  {Mart{\'{\i}}}  \& {M{\"u}ller}}{{Aloy} et~al.}{2000}]{2000ApJ...528L..85A}
{Aloy} M.-A.,  {G{\'o}mez} J.-L.,  {Ib{\'a}{\~n}ez} J.-M.,  {Mart{\'{\i}}}
  J.-M.,   {M{\"u}ller} E.,  2000, \mn@doi [\apjl] {10.1086/312436}, \href
  {http://adsabs.harvard.edu/abs/2000ApJ...528L..85A} {528, L85}

\bibitem[\protect\citeauthoryear{{Baldi}, {Capetti}  \& {Giovannini}}{{Baldi}
  et~al.}{2015}]{2015A&A...576A..38B}
{Baldi} R.~D.,  {Capetti} A.,   {Giovannini} G.,  2015, \mn@doi [\aap]
  {10.1051/0004-6361/201425426}, \href
  {http://adsabs.harvard.edu/abs/2015A%26A...576A..38B} {576, A38}

\bibitem[\protect\citeauthoryear{{Balmaverde}, {Baldi}  \&
  {Capetti}}{{Balmaverde} et~al.}{2008}]{2008A&A...486..119B}
{Balmaverde} B.,  {Baldi} R.~D.,   {Capetti} A.,  2008, \mn@doi [\aap]
  {10.1051/0004-6361:200809810}, \href
  {http://adsabs.harvard.edu/abs/2008A%26A...486..119B} {486, 119}

\bibitem[\protect\citeauthoryear{{Beck} \& {Krause}}{{Beck} \&
  {Krause}}{2005}]{2005AN....326..414B}
{Beck} R.,  {Krause} M.,  2005, \mn@doi [Astronomische Nachrichten]
  {10.1002/asna.200510366}, \href
  {http://adsabs.harvard.edu/abs/2005AN....326..414B} {326, 414}

\bibitem[\protect\citeauthoryear{{Blandford} \& {Rees}}{{Blandford} \&
  {Rees}}{1974}]{1974MNRAS.169..395B}
{Blandford} R.~D.,  {Rees} M.~J.,  1974, \mn@doi [\mnras]
  {10.1093/mnras/169.3.395}, \href
  {http://adsabs.harvard.edu/abs/1974MNRAS.169..395B} {169, 395}

\bibitem[\protect\citeauthoryear{{B{\"o}ttcher} \& {Dermer}}{{B{\"o}ttcher} \&
  {Dermer}}{2010}]{2010ApJ...711..445B}
{B{\"o}ttcher} M.,  {Dermer} C.~D.,  2010, \mn@doi [\apj]
  {10.1088/0004-637X/711/1/445}, \href
  {http://adsabs.harvard.edu/abs/2010ApJ...711..445B} {711, 445}

\bibitem[\protect\citeauthoryear{B\"{o}ttcher, Harris  \&
  Krawczynski}{B\"{o}ttcher et~al.}{2012}]{boettcher2012relativistic}
B\"{o}ttcher M.,  Harris D.,   Krawczynski H.,  2012, Relativistic Jets from
  Active Galactic Nuclei.
Wiley, Weinheim, Germany

\bibitem[\protect\citeauthoryear{Dermer \& Menon}{Dermer \&
  Menon}{2009}]{dermer2009high}
Dermer C.,  Menon G.,  2009, High Energy Radiation from Black Holes: Gamma
  Rays, Cosmic Rays, and Neutrinos.
Princeton Series in Astrophysics, Princeton University Press

\bibitem[\protect\citeauthoryear{{Falle}}{{Falle}}{1991}]{1991MNRAS.250..581F}
{Falle} S.~A.~E.~G.,  1991, \mn@doi [\mnras] {10.1093/mnras/250.3.581}, \href
  {http://adsabs.harvard.edu/abs/1991MNRAS.250..581F} {250, 581}

\bibitem[\protect\citeauthoryear{{Fanaroff} \& {Riley}}{{Fanaroff} \&
  {Riley}}{1974}]{1974MNRAS.167P..31F}
{Fanaroff} B.~L.,  {Riley} J.~M.,  1974, \mn@doi [\mnras]
  {10.1093/mnras/167.1.31P}, \href
  {http://adsabs.harvard.edu/abs/1974MNRAS.167P..31F} {167, 31P}

\bibitem[\protect\citeauthoryear{{Fromm}, {Perucho}, {Mimica}  \&
  {Ros}}{{Fromm} et~al.}{2016}]{2016A&A...588A.101F}
{Fromm} C.~M.,  {Perucho} M.,  {Mimica} P.,   {Ros} E.,  2016, \mn@doi [\aap]
  {10.1051/0004-6361/201527139}, \href
  {http://adsabs.harvard.edu/abs/2016A%26A...588A.101F} {588, A101}

\bibitem[\protect\citeauthoryear{{Gomez}, {Marti}, {Marscher}, {Ibanez}  \&
  {Marcaide}}{{Gomez} et~al.}{1995}]{1995ApJ...449L..19G}
{Gomez} J.~L.,  {Marti} J.~M.~A.,  {Marscher} A.~P.,  {Ibanez} J.~M.~A.,
  {Marcaide} J.~M.,  1995, \mn@doi [\apjl] {10.1086/309623}, \href
  {http://adsabs.harvard.edu/abs/1995ApJ...449L..19G} {449, L19}

\bibitem[\protect\citeauthoryear{{G{\'o}mez}, {Mart{\'{\i}}}, {Marscher},
  {Ib{\'a}{\~n}ez}  \& {Alberdi}}{{G{\'o}mez}
  et~al.}{1997}]{1997ApJ...482L..33G}
{G{\'o}mez} J.~L.,  {Mart{\'{\i}}} J.~M.,  {Marscher} A.~P.,  {Ib{\'a}{\~n}ez}
  J.~M.,   {Alberdi} A.,  1997, \mn@doi [\apjl] {10.1086/310671}, \href
  {http://adsabs.harvard.edu/abs/1997ApJ...482L..33G} {482, L33}

\bibitem[\protect\citeauthoryear{{Harris} \& {Krawczynski}}{{Harris} \&
  {Krawczynski}}{2006}]{2006ARA&A..44..463H}
{Harris} D.~E.,  {Krawczynski} H.,  2006, \mn@doi [\araa]
  {10.1146/annurev.astro.44.051905.092446}, \href
  {http://adsabs.harvard.edu/abs/2006ARA%26A..44..463H} {44, 463}

\bibitem[\protect\citeauthoryear{{Jones}, {Ryu}  \& {Engel}}{{Jones}
  et~al.}{1999}]{1999ApJ...512..105J}
{Jones} T.~W.,  {Ryu} D.,   {Engel} A.,  1999, \mn@doi [\apj] {10.1086/306772},
  \href {http://adsabs.harvard.edu/abs/1999ApJ...512..105J} {512, 105}

\bibitem[\protect\citeauthoryear{{Kaiser} \& {Alexander}}{{Kaiser} \&
  {Alexander}}{1997}]{1997MNRAS.286..215K}
{Kaiser} C.~R.,  {Alexander} P.,  1997, \mn@doi [\mnras]
  {10.1093/mnras/286.1.215}, \href
  {http://adsabs.harvard.edu/abs/1997MNRAS.286..215K} {286, 215}

\bibitem[\protect\citeauthoryear{{Laing} \& {Bridle}}{{Laing} \&
  {Bridle}}{2002}]{2002MNRAS.336.1161L}
{Laing} R.~A.,  {Bridle} A.~H.,  2002, \mn@doi [\mnras]
  {10.1046/j.1365-8711.2002.05873.x}, \href
  {http://adsabs.harvard.edu/abs/2002MNRAS.336.1161L} {336, 1161}

\bibitem[\protect\citeauthoryear{{Leismann}, {Ant{\'o}n}, {Aloy}, {M{\"u}ller},
  {Mart{\'{\i}}}, {Miralles}  \& {Ib{\'a}{\~n}ez}}{{Leismann}
  et~al.}{2005}]{2005A&A...436..503L}
{Leismann} T.,  {Ant{\'o}n} L.,  {Aloy} M.~A.,  {M{\"u}ller} E.,
  {Mart{\'{\i}}} J.~M.,  {Miralles} J.~A.,   {Ib{\'a}{\~n}ez} J.~M.,  2005,
  \mn@doi [\aap] {10.1051/0004-6361:20042520}, \href
  {http://adsabs.harvard.edu/abs/2005A%26A...436..503L} {436, 503}

\bibitem[\protect\citeauthoryear{{Mart{\'{\i}}}, {M{\"u}ller}, {Font},
  {Ib{\'a}{\~n}ez}  \& {Marquina}}{{Mart{\'{\i}}}
  et~al.}{1997}]{1997ApJ...479..151M}
{Mart{\'{\i}}} J.~M.,  {M{\"u}ller} E.,  {Font} J.~A.,  {Ib{\'a}{\~n}ez}
  J.~M.~Z.,   {Marquina} A.,  1997, \apj, \href
  {http://adsabs.harvard.edu/abs/1997ApJ...479..151M} {479, 151}

\bibitem[\protect\citeauthoryear{{Massaglia}, {Bodo}, {Rossi}, {Capetti}  \&
  {Mignone}}{{Massaglia} et~al.}{2016}]{2016A&A...596A..12M}
{Massaglia} S.,  {Bodo} G.,  {Rossi} P.,  {Capetti} S.,   {Mignone} A.,  2016,
  \mn@doi [\aap] {10.1051/0004-6361/201629375}, \href
  {http://adsabs.harvard.edu/abs/2016A%26A...596A..12M} {596, A12}

\bibitem[\protect\citeauthoryear{{Mignone} \& {McKinney}}{{Mignone} \&
  {McKinney}}{2007}]{2007MNRAS.378.1118M}
{Mignone} A.,  {McKinney} J.~C.,  2007, \mn@doi [\mnras]
  {10.1111/j.1365-2966.2007.11849.x}, \href
  {http://adsabs.harvard.edu/abs/2007MNRAS.378.1118M} {378, 1118}

\bibitem[\protect\citeauthoryear{{Mignone}, {Plewa}  \& {Bodo}}{{Mignone}
  et~al.}{2005}]{2005ApJS..160..199M}
{Mignone} A.,  {Plewa} T.,   {Bodo} G.,  2005, \mn@doi [\apjs]
  {10.1086/430905}, \href {http://adsabs.harvard.edu/abs/2005ApJS..160..199M}
  {160, 199}

\bibitem[\protect\citeauthoryear{{Mignone}, {Bodo}, {Massaglia}, {Matsakos},
  {Tesileanu}, {Zanni}  \& {Ferrari}}{{Mignone}
  et~al.}{2007}]{2007ApJS..170..228M}
{Mignone} A.,  {Bodo} G.,  {Massaglia} S.,  {Matsakos} T.,  {Tesileanu} O.,
  {Zanni} C.,   {Ferrari} A.,  2007, \mn@doi [\apjs] {10.1086/513316}, \href
  {http://adsabs.harvard.edu/abs/2007ApJS..170..228M} {170, 228}

\bibitem[\protect\citeauthoryear{{Mignone}, {Rossi}, {Bodo}, {Ferrari}  \&
  {Massaglia}}{{Mignone} et~al.}{2010}]{2010MNRAS.402....7M}
{Mignone} A.,  {Rossi} P.,  {Bodo} G.,  {Ferrari} A.,   {Massaglia} S.,  2010,
  \mn@doi [Mon. Not. R. Astron. Soc.] {10.1111/j.1365-2966.2009.15642.x}, \href
  {http://adsabs.harvard.edu/abs/2010MNRAS.402....7M} {402, 7}

\bibitem[\protect\citeauthoryear{{Mimica}, {Aloy}, {M{\"u}ller}  \&
  {Brinkmann}}{{Mimica} et~al.}{2004}]{2004A&A...418..947M}
{Mimica} P.,  {Aloy} M.~A.,  {M{\"u}ller} E.,   {Brinkmann} W.,  2004, \mn@doi
  [\aap] {10.1051/0004-6361:20034261}, \href
  {http://adsabs.harvard.edu/abs/2004A%26A...418..947M} {418, 947}

\bibitem[\protect\citeauthoryear{{Mimica}, {Aloy}  \& {M{\"u}ller}}{{Mimica}
  et~al.}{2007}]{2007A&A...466...93M}
{Mimica} P.,  {Aloy} M.~A.,   {M{\"u}ller} E.,  2007, \mn@doi [\aap]
  {10.1051/0004-6361:20066811}, \href
  {http://adsabs.harvard.edu/abs/2007A%26A...466...93M} {466, 93}

\bibitem[\protect\citeauthoryear{{Mimica}, {Aloy}, {Agudo}, {Mart{\'{\i}}},
  {G{\'o}mez}  \& {Miralles}}{{Mimica} et~al.}{2009}]{2009ApJ...696.1142M}
{Mimica} P.,  {Aloy} M.-A.,  {Agudo} I.,  {Mart{\'{\i}}} J.~M.,  {G{\'o}mez}
  J.~L.,   {Miralles} J.~A.,  2009, \mn@doi [\apj]
  {10.1088/0004-637X/696/2/1142}, \href
  {http://adsabs.harvard.edu/abs/2009ApJ...696.1142M} {696, 1142}

\bibitem[\protect\citeauthoryear{{Perucho} \& {Mart{\'{\i}}}}{{Perucho} \&
  {Mart{\'{\i}}}}{2007}]{2007MNRAS.382..526P}
{Perucho} M.,  {Mart{\'{\i}}} J.~M.,  2007, \mn@doi [\mnras]
  {10.1111/j.1365-2966.2007.12454.x}, \href
  {http://adsabs.harvard.edu/abs/2007MNRAS.382..526P} {382, 526}

\bibitem[\protect\citeauthoryear{{Posacki}, {Pellegrini}  \&
  {Ciotti}}{{Posacki} et~al.}{2013}]{2013MNRAS.433.2259P}
{Posacki} S.,  {Pellegrini} S.,   {Ciotti} L.,  2013, \mn@doi [\mnras]
  {10.1093/mnras/stt898}, \href
  {http://adsabs.harvard.edu/abs/2013MNRAS.433.2259P} {433, 2259}

\bibitem[\protect\citeauthoryear{{Rossi}, {Mignone}, {Bodo}, {Massaglia}  \&
  {Ferrari}}{{Rossi} et~al.}{2008}]{2008A&A...488..795R}
{Rossi} P.,  {Mignone} A.,  {Bodo} G.,  {Massaglia} S.,   {Ferrari} A.,  2008,
  \mn@doi [\aap] {10.1051/0004-6361:200809687}, \href
  {http://adsabs.harvard.edu/abs/2008A%26A...488..795R} {488, 795}

\bibitem[\protect\citeauthoryear{{Rueda-Becerril}, {Mimica}  \&
  {Aloy}}{{Rueda-Becerril} et~al.}{2014}]{2014MNRAS.438.1856R}
{Rueda-Becerril} J.~M.,  {Mimica} P.,   {Aloy} M.~A.,  2014, \mn@doi [\mnras]
  {10.1093/mnras/stt2335}, \href
  {http://adsabs.harvard.edu/abs/2014MNRAS.438.1856R} {438, 1856}

\bibitem[\protect\citeauthoryear{Rybicki \& Lightman}{Rybicki \&
  Lightman}{1979}]{rybicki1979radiative}
Rybicki G.,  Lightman A.,  1979, Radiative Processes in Astrophysics.
A Wiley-Interscience publication, Wiley

\bibitem[\protect\citeauthoryear{{Scheck}, {Aloy}, {Mart{\'{\i}}}, {G{\'o}mez}
  \& {M{\"u}ller}}{{Scheck} et~al.}{2002}]{2002MNRAS.331..615S}
{Scheck} L.,  {Aloy} M.~A.,  {Mart{\'{\i}}} J.~M.,  {G{\'o}mez} J.~L.,
  {M{\"u}ller} E.,  2002, \mn@doi [Mon. Not. R. Astron. Soc.]
  {10.1046/j.1365-8711.2002.05210.x}, \href
  {http://adsabs.harvard.edu/abs/2002MNRAS.331..615S} {331, 615}

\bibitem[\protect\citeauthoryear{{Sikora}, {Begelman}, {Madejski}  \&
  {Lasota}}{{Sikora} et~al.}{2005}]{2005ApJ...625...72S}
{Sikora} M.,  {Begelman} M.~C.,  {Madejski} G.~M.,   {Lasota} J.-P.,  2005,
  \mn@doi [\apj] {10.1086/429314}, \href
  {http://adsabs.harvard.edu/abs/2005ApJ...625...72S} {625, 72}

\makeatother
\end{thebibliography}







\bsp	
\label{lastpage}
\end{document}